\documentclass[twocolumn]{aastex701}

\usepackage{graphicx}
\usepackage{txfonts}

\begin{document}

\title{High-redshift Merger-induced Bar-like Galaxies in IllustrisTNG}

\author[orcid=0000-0001-7138-8899,gname=Ewa,sname=Lokas]{Ewa L. {\L}okas}

\affiliation{Nicolaus Copernicus Astronomical Center, Polish Academy of Sciences,
Bartycka 18, 00-716 Warsaw, Poland}

\email[show]{lokas@camk.edu.pl}

\shorttitle{High-redshift Merger-induced Bar-like Galaxies in IllustrisTNG}
\shortauthors{Ewa L. {\L}okas}

%\received{February 1, 2025}
%\revised{March 1, 2025}
%\accepted{\today}

\begin{abstract}
Recent discoveries made with JWST observations include a significant number of barred galaxies at high redshift. Their
origin remains unclear and their presence seems difficult to reproduce in cosmological simulations of galaxy formation
and evolution. In this Letter I present four examples of high-redshift bars selected from a sample of bar-like galaxies
studied previously using IllustrisTNG simulations. All the galaxies formed their bars at redshifts $z > 3$ via mergers
with smaller satellites, although one had its bar formed even earlier, at $z > 5$. The bars were born long, with
lengths on the order of 3 kpc, and grew in time. Three of the four galaxies were later accreted by clusters and
underwent multiple interactions with their respective brightest cluster galaxies. Their bar strength was to some extent
affected by these interactions but all the galaxies preserved their bar-like shape until the present time. By the end
of the evolution, all the galaxies lost their gas and stopped forming stars, they retained essentially no disk
component and were no longer rotationally supported. The examples demonstrate that high-$z$ bars do not evolve into
present-day barred disk galaxies similar to the Milky Way but rather into S0s or ellipticals typically found in galaxy
clusters.
\end{abstract}


%We could remove the \uat commands to avoid latex warning, just keep the text of keywords

\section{Introduction}

Bars are a common feature among late-type galaxies in the local Universe \citep{Buta2015}, where a majority of them
are classified as barred. The frequency of bars among higher-redshift galaxies is much less clear. Recently, our
knowledge in this area has improved substantially due to new discoveries of JWST. These observations
revealed well-developed bars in galaxies at high redshifts \citep{Guo2023, Costantin2023, Amvrosiadis2025} and allowed
for determinations of the bar fraction in the Universe at early times. Although the bar fraction was found to decrease
strongly with redshift, it turned out to be non-negligible even at $z = 3$ \citep{LeConte2024, Guo2025, Geron2025}. In
addition, the bars observed at high redshifts, for example the bar at $z = 3$ discovered by \citet{Costantin2023} with
JWST or the barred spiral galaxy found at $z = 4.4$ by \citet{Tsukui2024} using ALMA, appear to be quite mature, with
lengths on the order of 3 kpc.

Theoretical attempts to model these trends were not very successful so far. Cosmological simulations have persistently
failed to reproduce the observed behavior of bar fraction with redshift \citep{Peschken2019, Reddish2022}. Neither were
they able to produce long enough bars at early times. For example, using IllustrisTNG50 simulations \citet{Rosas2022}
managed to find bars in disks as early as at $z = 4$, but they were all of sub-kiloparsec length. The same was true for
the high-redshift bars studied by \citet{Bi2022} in their zoom-in simulations of galactic disks.

In their study of the 3 kpc-long bar observed at $z = 3$ with JWST \citet{Costantin2023} suggested that the object
could be a progenitor of a galaxy similar to the present-day Milky Way. However, such an object would not have to
evolve into anything similar to the Milky Way by the present time. In fact, studies of the evolution of
the Milky Way-M31 analogues from TNG50 \citep{Pillepich2024} and similar galaxies reveal that their bars grew mainly by
disk instability from small elongations at early times \citep{Rosas2022}, unlike the mature bars observed by JWST.
This suggests that the scenario for the formation of long bars at high redshifts may be entirely different and may
involve strong interactions with neighboring galaxies. Such an alternative way of forming bars in galaxies has been
extensively studied in the literature \citep{Noguchi1987, Gerin1990, Berentzen2004, Lokas2014, Lokas2016, Lokas2018}.
These studies demonstrated that bars can be efficiently induced in disk galaxies in flyby interactions with massive
enough perturbers, with small enough pericenters and preferably prograde orbits. Another interaction-related
possibility to form a bar is to have the disk perturbed as a result of mergers \citep{Cavanagh2020, Zhou2025}. Although
these mechanisms could operate at all times, they seem to be particularly promising at high redshifts since both flybys
and mergers were much more frequent in the early stages of galaxy evolution.

In this Letter I present four examples of simulated galaxies forming their mature bars at redshifts $z > 3$. They
were selected from the sample of bar-like galaxies from the IllustrisTNG project studied in \cite{Lokas2021}. These
galaxies were identified at $z = 0$ among systems with massive enough stellar component and a prolate shape. After
analyzing the evolutionary histories of these galaxies I divided them into three classes depending on their formation
scenarios and properties. Class A included objects whose bar-like structures had been induced by
an interaction with a more massive companion, typically a brightest cluster galaxy (BCG). Class B galaxies had their bar
induced by mergers or interactions with smaller satellites, or just by disk instability, and only later interacted with
more massive companions losing mass. The galaxies of class C had their bars formed by the same channels as those
of class B, but never interacted strongly with massive neighbors, as evidenced by their continuously growing mass. In a
recent paper \citep{Lokas2025a} I described one galaxy of class A that had the earliest bar formation time of this
class, $t = 2.2$ Gyr ($z = 2.9$). Here I discuss four further examples of early bars formed via interactions even
earlier and belonging to classes B and C.

\section{Formation and evolution of the bars}

The sample of bar-like galaxies described in \cite{Lokas2021} was selected from the last output (corresponding to the
present time, $z = 0$) of the TNG100-1 run of the IllustrisTNG suite of cosmological simulations \citep{Springel2018,
Marinacci2018, Naiman2018, Nelson2018, Pillepich2018}. The simulations follow the evolution of dark matter particles
and baryons in a box of size 100 Mpc, solving for gravity and hydrodynamics and applying additional sub-grid
prescriptions to describe processes like star formation and feedback. The selection was made from 6507 well
resolved galaxies, with total stellar masses $M_* > 10^{10}$ M$_\odot$, which corresponds to about $10^4$ stellar
particles per object. This choice ensures that the selected sample contains a large enough number of galaxies
for statistical analysis and at the same time is marginally sufficient for the morphological analysis including
the properties of the bars \citep{Dubinski2009}. The galaxies were then required to be strongly prolate, that is
possess the intermediate-to-long axis ratio of the stellar component $b/a < 0.6$. Among the 277 bar-like galaxies
identified in this way, 77 were found to be formed via a tidal interaction with a more massive companion (class A).
This tidally induced subsample was described in more detail in \citet{Lokas2025b} and includes the oldest tidally
induced bar-like galaxy presented in \citet{Lokas2025a}. The subsamples belonging to class B and C contained 74 and 126
galaxies, respectively. These two latter subsamples include four galaxies with bar formation times $z > 3$, which are
presented here.

The properties of the galaxies of interest were extracted from the publicly available simulation data described by
\citet{Nelson2019}. Here and in the following, the galaxies are referred to by their subhalo identification numbers (ID)
at $z=0$ unless stated otherwise. The catalogs provided with the data include the properties that can be used to
characterize the global shape of the galaxies, such as the axis ratios $b/a$ (intermediate-to-longest) and $c/a$
(shortest-to-longest) calculated from the eigenvalues of the mass tensor of the stellar component within two
stellar half-mass radii, $2 r_{1/2}$, which contains most of the galaxy stellar mass. A useful combination of those is
the triaxiality parameter $T = [1-(b/a)^2]/[1-(c/a)^2]$, which is low (close to zero) for oblate and high (close to
unity) for prolate systems. An additional quantity useful in the morphological characterization of galaxies is the
rotation parameter $f$ related to the fraction of stars on circular orbits, which measures the amount of rotational
support in the system \citep{Genel2015}, with $f > 0.4$ corresponding to well-developed disks \citep{Joshi2020}. In
addition to these, I calculated the commonly used measure of the bar strength \citep{Athanassoula2002} in the form of
the $m=2$ mode of the Fourier decomposition of the surface density distribution of stellar particles projected along
the short axis. It is given by $A_2 (R) = | \Sigma_j m_j \exp(2 i \theta_j) |/\Sigma_j m_j$, where $\theta_j$ is the
azimuthal angle of the $j$th star, $m_j$ is its mass, and the sum is over all particles in a given radial bin. The
single-value measurements presented here were obtained using all stars within two stellar half-mass radii, $2 r_{1/2}$.

\begin{figure*}
\centering
\includegraphics[width=15cm]{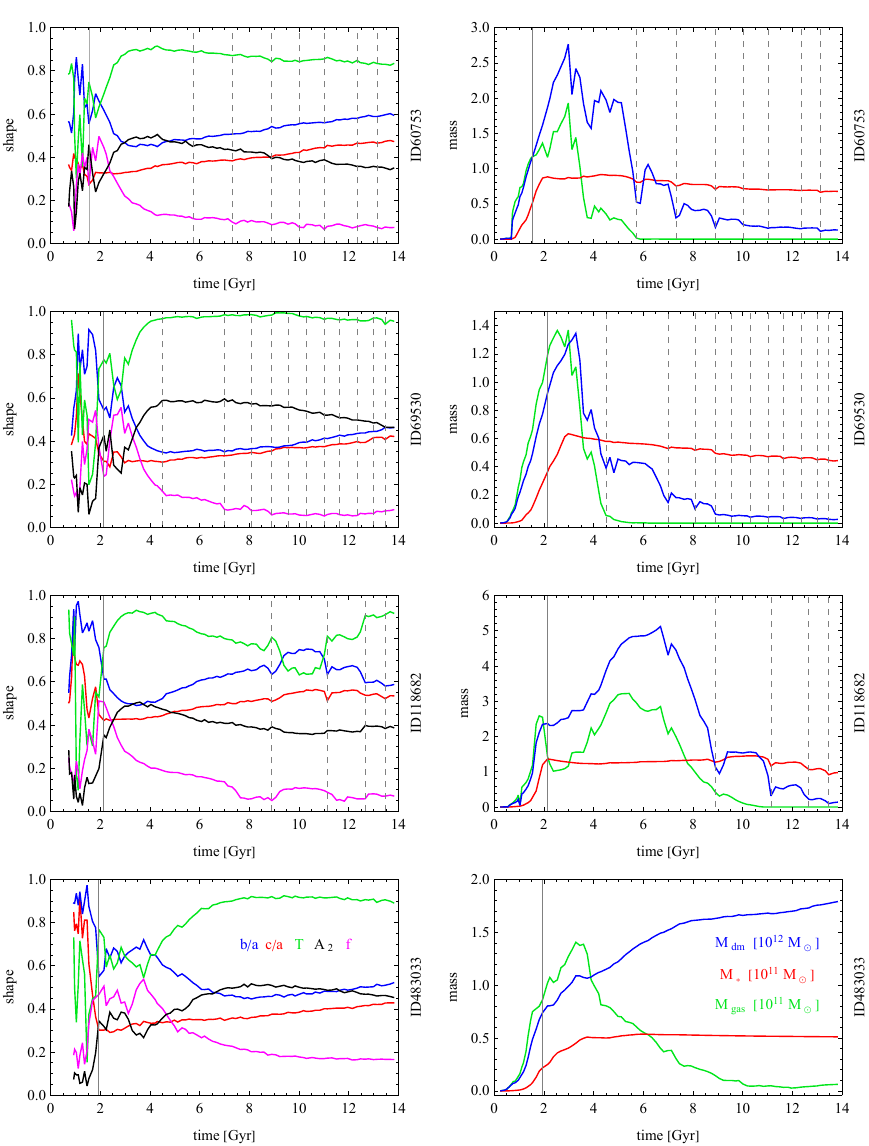}
\caption{Evolution of different measures of shape (left column) and mass (right column) over time for the four
bar-like galaxies. In the left-column panels the lines show the axis ratios $b/a$ (blue) and $c/a$ (red), the
triaxiality parameter $T$ (green), the bar mode $A_2$ (black), and the rotation parameter $f$ (magenta) of the stellar
component. In the right-column panels the lines show the total galaxy mass in dark matter (blue), stars (red) and gas
(green). The dark masses are given in units of $10^{12}$ M$_\odot$, while the stellar and gas masses are in
units of $10^{11}$ M$_\odot$. Vertical solid lines in all panels indicate the selected times $t_{\rm sel}$ at or after
bar formation. Vertical dashed lines in the upper six panels indicate pericenter passages around a cluster-size object.
The galaxies are identified by their subhalo ID numbers at $z = 0$ given on the right of each panel.}

\label{shapemass}
\end{figure*}

The evolution of these different measures of morphology in time is shown with different colors in the left-column
panels of Fig.~\ref{shapemass} for the four selected galaxies with IDs marked on the right side of each panel. The IDs
at $z = 0$ are also given in the first column of Table~\ref{properties}. The bar formation time is defined as the
output when the bar mode $A_2 (< 2 r_{1/2})$ (black lines) crosses the threshold of 0.2 and remains above it until the
end of the evolution. The redshift and the corresponding time of bar formation ($z_{\rm bf}$ and $t_{\rm bf}$) are
given in the second and third columns of the Table. As can be seen in the left-column panels of Fig.~\ref{shapemass},
the increase of $A_2$ is accompanied by a decrease in the axis ratio $b/a$ and an increase of triaxiality $T$, which
also signify the morphological transformation of a disk into a bar. As the bar becomes stronger later on, the
amount of rotational support decreases, as more radial orbits characteristic of the bar start to dominate. This
evolution is illustrated well by the parameter $f$, which drops from the initial high values characteristic of disks,
$f > 0.4$, down to values below 0.2 typical for objects no longer supported by rotation.

\begin{deluxetable*}{cccccccccccc}
\tablewidth{0pt}
\tablecaption{Properties of the bar-like galaxies}  \label{properties}
\tablehead{
\colhead{ID ($z=0$)} & \colhead{$z_{\rm bf}$} & \colhead{$t_{\rm bf}$} & \colhead{$z_{\rm sel}$} & \colhead{$t_{\rm
sel}$}& \colhead{ID ($z_{\rm sel}$)} & \colhead{$r_{1/2} (z_{\rm sel})$} & \colhead{$r_{\rm bar}
(z_{\rm sel})$} & \colhead{$A_{2, {\rm max}} (z_{\rm sel})$} & \colhead{$M_{\rm dm} (z_{\rm sel})$} &
\colhead{$M_{\rm gas} (z_{\rm sel})$} & \colhead{$M_* (z_{\rm sel})$}
\\
\colhead{} & \colhead{} & \colhead{(Gyr)} & \colhead{} & \colhead{(Gyr)} & \colhead{} & \colhead{(kpc)}
& \colhead{(kpc)} & \colhead{} & \colhead{($10^{12}$ M$_\odot$)} & \colhead{($10^{11}$ M$_\odot$)} &
\colhead{($10^{11}$ M$_\odot$)} }
\colnumbers
\startdata
60753 \  & 5.2 & 1.1 & 4.0 & 1.5 & 12682 & 1.02 & 2.2 & 0.59 & 1.2 & 1.2 & 0.5
\\
69530 \  & 3.3 & 1.9 & 3.0 & 2.1 & 48567 & 1.06 & 2.7 & 0.57 & 0.9 & 1.2 & 0.4
\\
118682 \ & 3.0 & 2.1 & 3.0 & 2.1 & 20998 & 1.17 & 3.2 & 0.53 & 2.4 & 1.4 & 1.4
\\
483033 \ & 3.3 & 1.9 & 3.3 & 1.9 & 49698 & 0.99 & 2.7 & 0.56 & 0.7 & 0.9 & 0.2
\\
\enddata
%\tablecomments{}
\end{deluxetable*}

\begin{figure*}
\centering
\includegraphics[width=16cm]{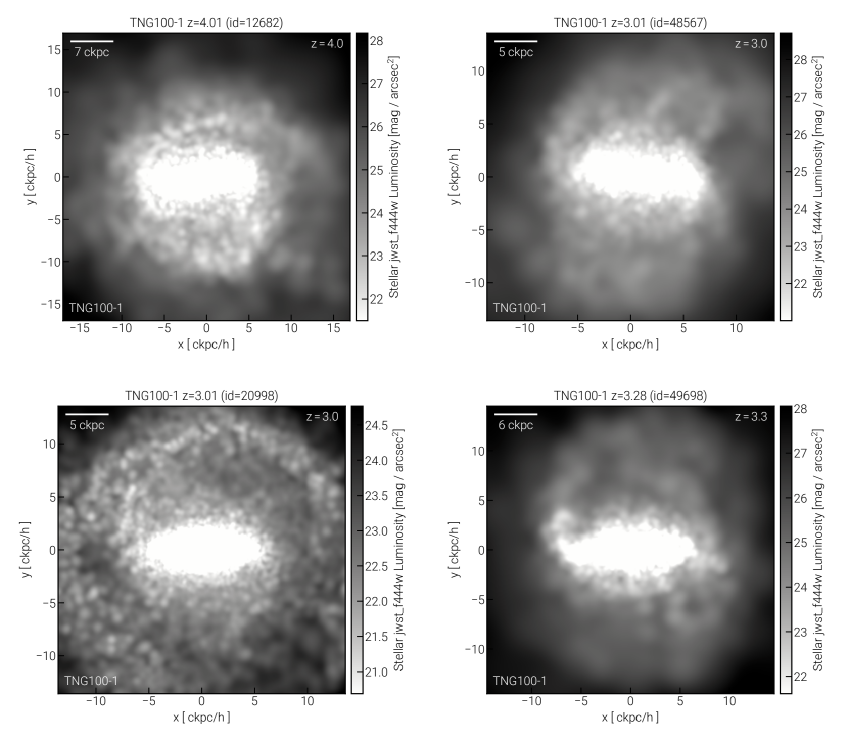}
\caption{Face-on images of the bar-like galaxies at redshift $z_{\rm sel}$, as they would be observed in the F444W
filter of JWST. The galaxies are identified by their IDs at $z_{\rm sel}$ given at the top of each panel and listed in
the sixth column of Table~\ref{properties}.}
\label{images}
\end{figure*}

The bars of the first two galaxies in Fig.~\ref{shapemass} are strongly variable initially so for a more detailed
analysis I selected slightly later outputs of $z_{\rm sel}$ (for the remaining two objects, $z_{\rm bf} = z_{\rm
sel}$). Their values, together with the corresponding times, $t_{\rm sel}$, are given in the fourth and fifth columns
of Table~\ref{properties}, while the IDs of the galaxies at this time are provided in the sixth column. The
seventh column of the Table contains the values of the stellar half-mass radius, $r_{1/2}$, of each galaxy at $z_{\rm
sel}$. The times $t_{\rm sel}$ are marked with vertical solid lines in all panels of Fig.~\ref{shapemass}.

Figure~\ref{images} shows the face-on views of the galaxies at these times as they would be observed in the F444W
filter of JWST. These images were created using the visualization tools provided with the IllustrisTNG data
release, which use stellar population models to translate the properties of stellar particles into the galaxy
luminosity, as described by \citet{Nelson2019}. They do not include any instrument-specific effects such as PSF or
noise. The bars are well visible in these images, as the most notable morphological features, and in some of the
objects are accompanied by much less pronounced spiral arms.

\begin{figure}
\centering
\includegraphics[width=7.3cm]{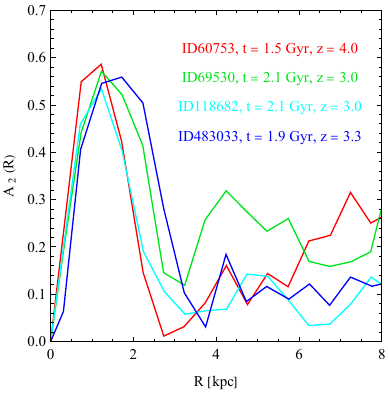}
\caption{Profiles of bar mode, $A_2 (R),$ for the bar-like galaxies at selected times. Measurements were
carried out in bins of $\Delta R = 0.5$ kpc.}
\label{a2profiles}
\end{figure}

The properties of the bars can be characterized in more detail by calculating the profiles $A_2 (R)$ as a function of
the cylindrical radius, rather than single measurements. These are shown in Fig.~\ref{a2profiles} for the selected
outputs at redshifts $z_{\rm sel}$, the same as in Fig.~\ref{images}. The profiles show a typical behavior,
characteristic of barred galaxies, first increasing with radius, reaching a maximum $A_{2, {\rm max}}$, and then
decreasing. The length of the bar can be estimated as a radius in the declining part of the $A_2 (R)$ profile, until
which the phase of the bar mode remains approximately constant. The bar lengths estimated in this way are given in the
eighth column of Table~\ref{properties} and the corresponding values of $A_{2, {\rm max}}$ in the ninth. One can see
that the lengths of the bars are considerable, of the order of 3 kpc, except for the first, earliest bar, which is a
little shorter. In particular, the bars are $2.2-2.7$ times longer than the stellar half-mass radii (seventh
column of Table~\ref{properties}), which means that they are much longer than the lengths of observed bars embedded in
disks at low redshifts, which typically have $r_{\rm bar}/r_{1/2} \sim 1$ \citep{Kim2021}. The bars are also quite
strong, with $A_{2, {\rm max}} > 0.5$, and this strength is similar in all four cases.

\begin{figure}
\centering
\hspace{0.8cm}
\includegraphics[width=3.5cm]{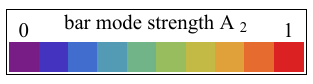}\\
\vspace{0.15cm}
\includegraphics[width=9cm]{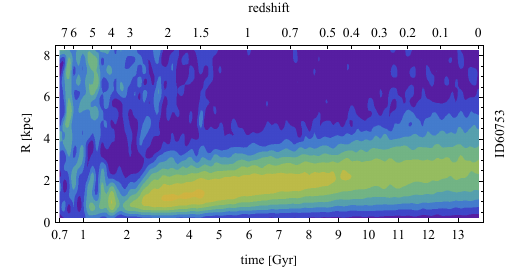} \\
\vspace{0.12cm}
\includegraphics[width=9cm]{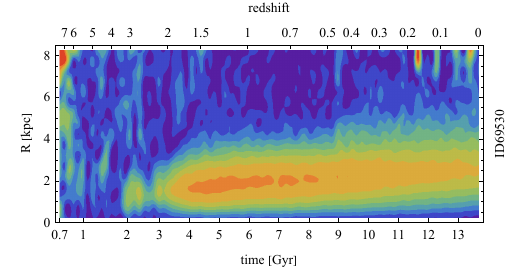} \\
\vspace{0.12cm}
\includegraphics[width=9cm]{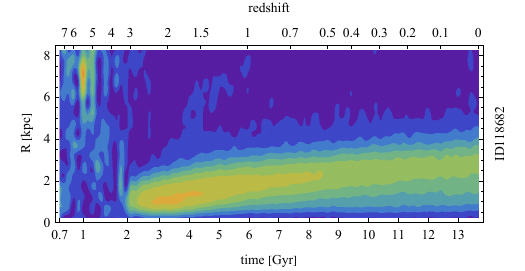} \\
\vspace{0.12cm}
\includegraphics[width=9cm]{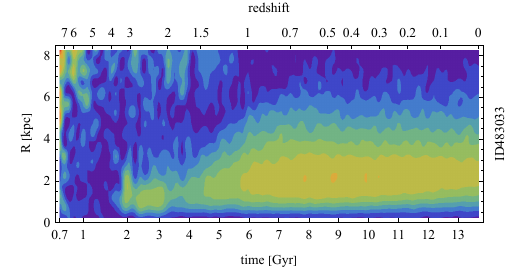}
\caption{Evolution of the profiles of the bar mode, $A_2 (R)$, in the four galaxies over time.
Measurements were carried out in bins of $\Delta R = 0.5$ kpc.
The galaxies are identified by their subhalo ID numbers at $z = 0$ given on the right of each panel.}
\label{a2modestime}
\end{figure}

The evolution of the bar mode in the selected galaxies can be seen in its entirety in Fig.~\ref{a2modestime}, where the
$A_2 (R)$ profiles are plotted as a function of time in the color-coded form. Plots of this kind can be used to
estimate both the strength of the bar (redder colors indicate stronger bars) and its length (where the strength
declines with radius). The plots demonstrate that in almost all cases the bars are initially strongly variable and they
are at their strongest a few gigayears after formation, but not at the end of the evolution. This is also reflected by
the single-value measurements of $A_2 (< 2 r_{1/2})$ shown with the black lines in the left-column panels of
Fig.~\ref{shapemass}, which tend to decrease in the later stages of the evolution.

Let us now have a closer look at the origin of the bars. Careful inspection of the neighborhood of the galaxies at the
time of bar formation reveals that they were all formed after experiencing mergers with nearby smaller
objects. In the case of ID60753, which has the earliest bar formation time of $t_{\rm bf} = 1.1$ Gyr ($z_{\rm bf} =
5.2$) the bar seems to be the result of four mergers with small satellites of mass above $10^9$ M$_\odot$ present in
its neighborhood at $t = 1$ Gyr, which interacted and merged with it at $t = 1 - 2$ Gyr. The galaxy also experienced
two other mergers with more massive satellites later on, at $t = 3.5$ Gyr and $t = 4.5$ Gyr, but they did not seem to
affect the already formed bar.

The galaxy ID69530 formed its bar at $t = 1.9$ Gyr ($z = 3.3$) when it merged with a satellite of mass larger
than $10^9$ M$_\odot$ and interacted with another. Two more mergers occurred until $t = 3.4$ Gyr, which is reflected in
the variability of the $A_2$ mode at early times, similar to the one seen in ID60753, after which period the bar
started to grow steadily. The case of ID118682 is special among the galaxies studied here in the way its bar formed: it
was induced very quickly, with $A_2$ growing strongly and monotonically, at $t = 2.1$ Gyr ($z = 3.0$) by mergers with
two particularly massive satellites, both with masses above $10^{10}$ M$_\odot$. Galaxy ID483033, on the other hand,
experienced multiple, most numerous mergers with objects in the mass range $10^{9-10}$ M$_\odot$ around the time of its
bar formation, $t = 1.9$ Gyr ($z = 3.3$), which resulted in some variability of the $A_2$ mode initially, similar to
the first two cases.

The evolution of the masses of the galaxies in different components (right-column panels of Fig.~\ref{shapemass})
reveals their later fates. Except for the last one (ID483033, lower-right panel), all experienced strong mass loss in
the dark matter component (blue lines) and gas (green lines). It turns out that during the later stages of the
evolution all three became cluster members and interacted with their respective BCGs. The
pericenter passages around the BCGs are marked with vertical dashed lines in the six upper panels of
Fig.~\ref{shapemass}. One can see that at the times when the bars have formed ($t_{\rm sel}$), they were all
already quite massive. The values of the masses of the galaxies in different components at that time are given in the
last three columns of Table~\ref{properties}.

The tidal stirring inherent in such interactions generally resulted in making the bars weaker, except for ID118682
where the bar variability was more complicated. As can be seen in the panel showing the shape evolution for this galaxy
in Fig.~\ref{shapemass}, the shape (in particular the triaxiality parameter $T$) changed abruptly during pericenters
(although the galaxy also had its last significant merger at the time of the first pericenter). This effect, described
in detail in \citet{Lokas2014}, is related to the orientation of the bar at the pericenter. The bar strength can
increase (decrease) if the torque of the tidal force acting on the bar slows down (speeds up) its rotation. It is also
worth noting that ID118682 has the largest stellar mass of all the galaxies studied here. It is due to significant gas
accretion and enhanced star formation resulting from mergers with particularly massive satellites at the time of bar
formation. On the other hand, ID483033 is the only galaxy of the four that evolved in isolation, outside of groups or
clusters, and therefore its dark mass grew monotonically; it is also the only one that retained some gas until the
present time. All the other three galaxies had their gas ram-pressure stripped during their evolution in the cluster
environment.

\section{Discussion}

I presented four examples of simulated bar-like galaxies from IllustrisTNG formed at $z > 3$ by mergers with smaller
satellites. The masses of the companions were typically of a few times $10^9$ M$_\odot$, in agreement with the findings
from the controlled simulations of \citet{Zhou2025}, who estimated such perturber masses to be sufficient to induce a
bar. Contrary to the case described previously in \citet{Lokas2025a}, these bars were not induced by an interaction
with a more massive galaxy, a progenitor of a BCG, although three of them became cluster members during their
subsequent evolution and had their bars modified by the cluster environment. All the bars presented here were born
long, with lengths on the order of 3 kpc, rather than being grown from small elongations as is characteristic of bars
forming by disk instability. All the bars were formed in gas-rich systems, with total gas mass typically a few times
larger, or at least comparable, to the total stellar mass. However, the gas fraction was always below 40\% of the
baryon content within two stellar half-mass radii, $2 r_{1/2}$, where the bar-like shape dominates, in agreement with
earlier studies, which found that lower gas fractions make galaxies more susceptible to bar formation
\citep{Shlosman1993, Athanassoula2013, Lokas2020}.

The objects presented here can be considered as simulated counterparts of high-redshift bars recently discovered with
JWST and ALMA. In particular, the lengths of the bars measured here are on the order of 3 kpc, similar to the bar at $z
= 3$ discovered by \citet{Costantin2023} and at $z = 4.4$ by \citet{Tsukui2024}. The galaxy ceers-2112 studied by
\citet{Costantin2023} has been speculated to be a progenitor of a Milky Way-like galaxy, given its bar length and
stellar mass. While similar to the observed galaxies at these early times, the simulated objects studied here did not
evolve into anything like the present-day Milky Way. By redshift $z = 0$ their stellar components are dominated by
a much longer bar-like shape, with hardly any disk and little rotational support, are completely or almost entirely
devoid of gas, do not form stars and are red. They are therefore more similar to S0s or ellipticals than a typical
spiral like the Milky Way. This suggests that early bars will probably not be found in simulations among progenitors of
present-day barred disks or Milky Way-M31 analogues \citep{Rosas2022, Pillepich2024}. Such galaxies typically form
their bars via dynamical instability of cold, massive disks and grow their bars slowly from small elongations.

The galaxies described in this work are the only ones with bar formation time $z > 3$ among the 277 bar-like galaxies
identified at $z = 0$ and studied in \citet{Lokas2021}. This means that these are bar-like objects that not only formed
early but also survived in this shape until the present. Relaxing the required time of bar formation a little and
adopting for example the threshold of $z > 2.5$ one finds twelve objects from the same total sample of bar-like
galaxies. This does not mean that these are the only bar-like objects that existed at these early redshifts in these
simulations. For example, selecting bar-like galaxies at $z = 3$ (rather than at $z = 0$) in the TNG100-1 run among 1358
galaxies with stellar mass above $10^{10}$ M$_\odot$ at that time one finds 49 objects satisfying the condition for the
axis ratio $b/a < 0.6$. Only some of these objects survived in a similar form until the present, while many merged
with bigger structures. Most of them, however, could be candidates for the analogues of the observed high-redshift bars,
since observers only have one snapshot of a given galaxy at their disposal and cannot predict its later evolution.

The evolutionary histories of the objects presented here provide a viable scenario for the formation of high-redshift
bars since mergers and tidal interactions occur more frequently at the early stages of galaxy evolution. In addition,
at early times the galactic disks tend to be hotter, more gas-rich and more turbulent, which makes the onset of the
inherent bar instability more problematic. This scenario is also supported by the observations that take into account
the environment of high-$z$ bars. In particular, many of the high-$z$ bars studied by \citet{Guo2023, Guo2025} appear
to have companions and the fraction of barred galaxies with companions turns out to be higher than for unbarred
ones.

\begin{acknowledgments}
I am grateful to the anonymous reviewer for useful comments and to the IllustrisTNG team for making their
simulations publicly available. Computations for this work have been performed using the computer cluster at the
Nicolaus Copernicus Astronomical Center of the Polish Academy of Sciences (CAMK PAN).
\end{acknowledgments}


\begin{thebibliography}{}

\bibitem[{A. Amvrosiadis et al.}(2025)]{Amvrosiadis2025} Amvrosiadis, A., Lange, S., Nightingale, J. W., et al.
        2025, MNRAS, 537, 1163
\bibitem[{E. Athanassoula \& A. Misiriotis}(2002)]{Athanassoula2002} Athanassoula, E., \& Misiriotis, A. 2002,
        MNRAS, 330, 35
\bibitem[{E. Athanassoula et al.}(2013)]{Athanassoula2013} Athanassoula, E., Machado, R. E. G., \& Rodionov, S. A. 2013,
        MNRAS, 429, 1949
\bibitem[{I. Berentzen et al.}(2004)]{Berentzen2004} Berentzen, I., Athanassoula, E., Heller, C. H., \& Fricke, K. J.
        2004, MNRAS, 347, 220
\bibitem[{D. Bi et al.}(2022)]{Bi2022} Bi, D., Shlosman, I., \& Romano-D\'{i}az, E. 2022, ApJ, 934, 52
\bibitem[{R. J. Buta et al.}(2015)]{Buta2015} Buta, R. J., Sheth, K., Athanassoula, E., et al. 2015, ApJS, 217, 32
\bibitem[{M. K. Cavanagh \& K. Bekki}(2020)]{Cavanagh2020} Cavanagh, M. K., \& Bekki, K. 2020, A\&A, 641, A77
\bibitem[{L. Costantin et al.}(2023)]{Costantin2023} Costantin, L., P\'{e}rez-Gonz\'{a}lez, P. G., Guo, Y., et al.
        2023, Nature, 623, 499
\bibitem[{J. Dubinski et al.}(2009)]{Dubinski2009} Dubinski, J., Berentzen, I., \& Shlosman, I. 2009, ApJ, 697, 293
\bibitem[{S. Genel et al.}(2015)]{Genel2015} Genel, S., Fall, S. M., Hernquist, L., et al. 2015, ApJ, 804, L40
\bibitem[{M. Gerin et al.}(1990)]{Gerin1990} Gerin, M., Combes, F., \& Athanassoula, E. 1990, A\&A, 230, 37
\bibitem[{T. Geron et al.}(2025)]{Geron2025} G\'{e}ron, T., Smethurst, R. J., Dickinson, H., et al. 2025, ApJ, 987, 74
\bibitem[{Y. Guo et al.}(2023)]{Guo2023} Guo, Y., Jogee, S., Finkelstein, S. L., et al. 2023, ApJ, 945, L10
\bibitem[{Y. Guo et al.}(2025)]{Guo2025} Guo, Y., Jogee, S., Wise, E., et al. 2025, ApJ, 985, 181
\bibitem[{G. D. Joshi et al.}(2020)]{Joshi2020} Joshi, G. D., Pillepich, A., Nelson, D., et al. 2020, MNRAS, 496, 2673
\bibitem[{T. Kim et al.}(2021)]{Kim2021} Kim, T., Athanassoula, E., \& Sheth, K., et al. 2021, ApJ, 922, 196
\bibitem[{Z. A. Le Conte et al.}(2024)]{LeConte2024} Le Conte, Z. A., Gadotti, D. A., Ferreira, L., et al. 2024,
        MNRAS, 530, 1984
\bibitem[{E. L. {\L}okas}(2018)]{Lokas2018} {\L}okas, E. L. 2018, ApJ, 857, 6
\bibitem[{E. L. {\L}okas}(2020)]{Lokas2020} {\L}okas, E. L. 2020, A\&A, 634, A122
\bibitem[{E. L. {\L}okas}(2021)]{Lokas2021} {\L}okas, E. L. 2021, A\&A, 647, A143
\bibitem[{E. L. {\L}okas}(2025a)]{Lokas2025a} {\L}okas, E. L. 2025a, A\&A, 700, A258
\bibitem[{E. L. {\L}okas}(2025b)]{Lokas2025b} {\L}okas, E. L. 2025b, A\&A, 702, A7
\bibitem[{E. L. {\L}okas et al.}(2014)]{Lokas2014} {\L}okas, E. L., Athanassoula, E., Debattista, V. P., et al. 2014,
        MNRAS, 445, 1339
\bibitem[{E. L. {\L}okas et al.}(2016)]{Lokas2016} {\L}okas, E. L., Ebrov\'{a}, I., del Pino, A., et al. 2016, ApJ, 826,
	227
\bibitem[{F. Marinacci et al.}(2018)]{Marinacci2018} Marinacci, F., Vogelsberger, M., Pakmor, R., et al. 2018,
        MNRAS, 480, 5113
\bibitem[{J. P. Naiman et al.}(2018)]{Naiman2018} Naiman, J. P., Pillepich, A., Springel, V., et al., 2018, MNRAS, 477,
	1206
\bibitem[{D. Nelson et al.}(2018)]{Nelson2018} Nelson, D., Pillepich, A., Springel, V., et al. 2018, MNRAS, 475, 624
\bibitem[{D. Nelson et al.}(2019)]{Nelson2019} Nelson, D., Springel, V., Pillepich, A., et al. 2019,
         Computat. Astroph. Cosmol., 6, 2
\bibitem[{M. Noguchi}(1987)]{Noguchi1987} Noguchi, M. 1987, MNRAS, 228, 635
\bibitem[{N. Peschken \& E. L. {\L}okas}(2019)]{Peschken2019} Peschken, N., \& {\L}okas, E. L. 2019, MNRAS, 483, 2721
\bibitem[{A. Pillepich et al.}(2018)]{Pillepich2018} Pillepich, A., Nelson, D., Hernquist, L., et al. 2018,
        MNRAS, 475, 648
\bibitem[{A. Pillepich et al.}(2024)]{Pillepich2024} Pillepich, A., Sotillo-Ramos, D., Ramesh, R., et al. 2024,
        MNRAS, 535, 1721
\bibitem[{J. Reddish et al.}(2022)]{Reddish2022} Reddish, J., Kraljic, K., Petersen, M. S., et al. 2022, MNRAS, 512, 160
\bibitem[{Y. Rosas-Guevara et al.}(2022)]{Rosas2022} Rosas-Guevara, Y., Bonoli, S., Dotti, M., et al. 2022, MNRAS, 512,
        5339
\bibitem[{I. Shlosman \& M. Noguchi }(1993)]{Shlosman1993} Shlosman, I., \& Noguchi, M. 1993, ApJ, 414, 474
\bibitem[{V. Springel et al.}(2018)]{Springel2018} Springel, V., Pakmor, R., Pillepich, A., et al. 2018, MNRAS, 475, 676
\bibitem[{T. Tsukui et al.}(2024)]{Tsukui2024} Tsukui, T., Wisnioski, E., Bland-Hawthorn, J., et al. 2024, MNRAS, 527,
	8941
\bibitem[{Y. F. Zhou et al.}(2025)]{Zhou2025} Zhou, Y. F., Li, Z., \& Li, H. 2025, ApJ, 988, 205


\end{thebibliography}
\end{document}